\documentclass[pra,twocolumn,aps,showpacs,superscriptaddress]{revtex4-1}

\usepackage{amsmath}
\usepackage{amssymb}
\usepackage{hyperref}
\usepackage{graphicx}
\usepackage{color}
\usepackage{bm}
\everymath{\displaystyle}
\begin{document}


\title{Goldstone modes and bifurcations in phase-separated binary condensates
       at finite temperature}

\author{Arko Roy}
\affiliation{Physical Research Laboratory,
             Navrangpura, Ahmedabad-380009, Gujarat,
             India}
\author{S. Gautam}
\affiliation{Department of Physics, Indian Institute of Science,
             Bangalore-560012, India
             }
\author{D. Angom}
\affiliation{Physical Research Laboratory,
             Navrangpura, Ahmedabad-380009, Gujarat,
             India}

\begin{abstract}
  We show that the third Goldstone mode, which emerges in binary condensates 
at phase-separation, persists to higher inter-species interaction for
density profiles where one component is surrounded on both sides by the other 
component. This is not the case with symmetry-broken density profiles
where one species is to entirely to the left and the other is entirely to
the right. We, then, use Hartree-Fock-Bogoliubov theory with Popov 
approximation to examine the mode evolution at $T\neq0$ and demonstrate the 
existence of mode bifurcation near the critical temperature. The Kohn mode, 
however, exhibits deviation from the natural frequency at finite temperatures 
after the phase separation. This is due to the exclusion of the non-condensate 
atoms in the dynamics. 
\end{abstract}



\maketitle


\section{Introduction}

  The remarkable feature of binary condensates or two-species Bose-Einstein 
condensates (TBECs) is the phenomenon of phase separation
\cite{ho_96,trippenbach_2000}. This relates the system to
novel phenomena in nonlinear dynamics and pattern formation, non-equilibrium 
statistical mechanics, optical systems and phase transitions in condensed 
matter systems. Experimentally, TBECs have been realized  in the mixture of 
two different alkali atoms \cite{modugno_02,lercher_11,mccarron_11}, and in 
two different isotopes \cite{inouye_98} and hyperfine states 
\cite{stamper_kurn_98,myatt_97} of an atom. Most importantly, in experiments, 
the TBEC can be steered from miscible to phase-separated domain or 
vice-versa \cite{papp_08,tojo_10} through a Feshbach resonance. 
These have motivated theoretical investigations on stationary states
\cite{ho_96,gautam-11},
dynamical instabilities \cite{sasaki_09,gautam_10,kadokura-12}
and collective excitations 
\cite{stringari_96,pu_98,graham_98,gordon_98,kuopanportii_12,ticknor_13,takeuchi_13}of TBECs. 

 In this paper, we report the development of Hartree-Fock-Bogoliubov theory
with Popov (HFB-Popov) approximation \cite{griffin_96} for TBECs. We use it to
investigate the evolution of Goldstone modes and mode energies as function of 
the inter-species interaction and temperature, respectively.  Recent works 
\cite{ticknor_13,takeuchi_13} reported  the existence of an additional 
Goldstone mode at phase separation in the symmetry-broken density profiles,
which we refer to as the {\em side-by-side} density profiles. We, 
however, demonstrate that in the other type of density profile where
one of the species is surrounded on both sides by the other,
which we refer to as the {\em sandwich} type, the mode 
evolves very differently.  To include the finite temperature effects, besides
HFB-Popov approximation, there are other different approaches. These include
projected Gross-Pitaevskii (GP) equation \cite{blakie_08}, stochastic GP 
equation(SGPE)\cite{proukakis_08} and Zaremba-Nikuni-Griffin(ZNG) formalism
\cite{zaremba-99}. For the present work we have chosen the HFB-Popov 
approximation, which is a gapless theory and satisfies the Hugenholtz-Pines 
theorem \cite{hugenholtz_59}. The method is particularly well suited to 
examine the evolution of the low-lying modes. It has been used extensively 
in single species BEC to study finite temperature effects to mode energies 
\cite{griffin_96,hutchinson_97,dodd_98,gies_04} and agrees well with the 
experimental results  \cite{jin_97} at low temperatures. In TBECs, the 
HFB-Popov approximation has been used in the miscible domain \cite{pires_08} 
and in this paper, we describe the results for the phase-separated domain. 
Other works which have examined the finite temperature effects in TBECs use 
Hartree-Fock treatment with or without trapping potential
\cite{ohberg_98,zhang_07} and semi-classical approach \cite{ohberg_99}. 
Although, HFB-Popov does have the advantage vis-a-vis calculation of the 
modes, it is nontrivial to get converged solutions. In the present work, we 
consider the TBEC of $^{87}$Rb-$^{133}$Cs \cite{lercher_11,mccarron_11}, which 
have widely differing $s$-wave scattering lengths and masses. This choice 
does add to the severity of the convergence issues but this also makes it a 
good test for the methods we use. We choose the parameter domain 
where the system is quasi-1D and a mean-field  description like HFB-Popov is 
applicable. The quasi-1D trapped bosons exhibit a rich phase structure as a 
function of density and interaction strengths \cite{petrov_00}. 
For comparison with the experimental results we also consider the parameters
as in the experiment \cite{mccarron_11}. We find that, like in Ref. 
\cite{egorov_13}, the quasi-1D description are in good agreement with the 
condensate density profiles of 3D calculations \cite{pattinson_13}.


\section{Theory}

For a highly anisotropic cigar shaped harmonic trapping 
potential $(1/2)m(\omega_x^2x^2 + \omega_y^2y^2 + \omega_z^2z^2)$, with
$\omega_x=\omega_y=\omega_{\perp} \gg \omega_z $. In this case, we can 
integrate out the condensate wave function along $xy$ and reduce it to a 
quasi-1D system. The transverse degrees of freedom are then frozen and the 
system is confined in the harmonic oscillator ground state along the 
transverse direction for which $\hbar\omega_{\perp}\gg \mu_k$. We thus 
consider excitations present only in the axial direction
$z$ \cite{mateo_08,mateo_09}. The grand-canonical Hamiltonian, in the 
second quantized form, describing the mixture of two interacting BECs is then
\begin{eqnarray}
  H &=& \sum_{k=1,2}\int dz\hat{\Psi}_{k}^\dagger(z,t)
        \bigg[-\frac{\hbar^{2}}{2m_k}\frac{\partial ^2}{\partial z^2} 
        + V_k(z)-\mu_k\nonumber\\ 
    & & + \frac{U_{kk}}{2}\hat{\Psi}_{k}^\dagger(z,t)\hat{\Psi}_{k}
        (z,t)\bigg]\hat{\Psi}_{k}(z,t)\nonumber\\ 
    & & + U_{12}\int dz
        \hat{\Psi}_{1}^\dagger(z,t) \hat{\Psi}_{2}^\dagger(z,t)
        \hat{\Psi}_{1}(z,t)\hat{\Psi}_{2}(z,t),
\label{hamiltonian} 
\end{eqnarray}
where $k=1,2$ is the species index, $\hat{\Psi}_k$'s are the Bose field 
operators of the two different species, and $\mu_k$'s  are the chemical 
potentials. The strength of intra and inter-species interactions are 
$U_{kk} = (a_{kk}\lambda)/m_{k}$ and $U_{12}=(a_{12}\lambda)/(2m_{12})$, 
respectively, where $\lambda = (\omega_{\perp}/\omega_z) \gg 1$ is the 
anisotropy parameter, $a_{kk}$ is the $s$-wave scattering length, $m_k$'s are 
the atomic masses of the species and $m_{12}=m_1 m_2/(m_1+m_2)$. In the present 
work we consider all the interactions are repulsive, that is 
$a_{kk},a_{12} > 0$. The equation of motion of the Bose field operators is 
\begin{equation}
 i\hbar\frac{\partial}{\partial t}
 \begin{pmatrix}
   \hat{\Psi}_1\\
   \hat{\Psi}_2
\end{pmatrix} \!\!
= \!\!\begin{pmatrix}
  \hat{h}_1 + U_{11}\hat{\Psi}_1^\dagger\hat{\Psi}_1 & U_{12}
           \hat{\Psi}_2^\dagger \hat{\Psi}_1\\
   U_{12}\hat{\Psi}_1^\dagger\hat{\Psi}_2             & \hat{h}_2
           + U_{22}\hat{\Psi}_2^\dagger \hat{\Psi}_2
  \end{pmatrix} \!\!\!
  \begin{pmatrix}
    \hat{\Psi}_1\\
    \hat{\Psi}_2  
  \end{pmatrix} \nonumber
\label{twocomp}
\end{equation}
where $\hat{h}_{k}= (-\hbar^{2}/2m_k)\partial ^2/\partial z^2 +V_k(z)-\mu_k$.
For compact notations, we refrain from writing the explicit
dependence of $\hat{\Psi}_k$ on $z$ and $t$. Since a majority of the atoms 
reside in the ground state for the temperature regime relevant to the 
experiments ($T\leqslant 0.65T_c$ ) \cite{dodd_98}, the condensate part
can be separated out from the Bose field operator $\hat{\Psi}({\mathbf r},t)$. 
The non-condensed or the thermal cloud of atoms are then the fluctuations of
the condensate field. Here, $T_c$ is the critical temperature of ideal
gas in a harmonic confining potential. Accordingly, we define 
\cite{griffin_96}, $\hat{\Psi}(z,t) = \Phi (z) + \tilde\Psi(z,t)$, 
where $\Phi(z)$  is a $c$-field and represents the condensate, and 
$\tilde\Psi(z,t) $ is the fluctuation part. In two component 
representation
\begin{eqnarray}
\begin{pmatrix}
 \hat{\Psi}_1\\
 \hat{\Psi}_2
\end{pmatrix}
 =
\begin{pmatrix}
 \phi_1\\
 \phi_2
\end{pmatrix}
 +
\begin{pmatrix}
\tilde\psi _1\\
\tilde\psi _2\\
\end{pmatrix},
\end{eqnarray}
where $\phi_k(z)$ and $\tilde{\psi}_k(z)$ are the condensate and fluctuation 
part of the $k$th species. Thus for a TBEC, $\phi_k$s are the static 
solutions of the coupled generalized GP equations, with time-independent 
HFB-Popov approximation, given by 
\begin{subequations}
\begin{eqnarray}
  \hat{h}_1\phi_1 + U_{11}\left[n_{c1}+2\tilde{n}_{1}\right]\phi_1
  +U_{12}n_2\phi_1=0,\\
  \hat{h}_2\phi_2 + U_{22}\left[n_{c2}+2\tilde{n}_{2}\right]\phi_2
  +U_{12}n_1\phi_2=0,
\end{eqnarray}
\label{gpe}
\end{subequations}
where, $n_{ck}(z)\equiv|\phi_k(z)|^2$,
$\tilde{n}_k(z)\equiv\langle\tilde{\psi}_{k}^{\dagger}(z,t)
\tilde{\psi}_k(z,t)\rangle$, and $n_k(z) = n_{ck}(z)+ \tilde{n}_k(z)$
are the local condensate, non-condensate, and total density,
respectively. Using Bogoliubov transformation
\begin{equation}
   \tilde{\psi}_k(z,t) =\sum_{j}\left[u_{kj}(z)
      \hat{\alpha}_j(z) e^{-iE_{j}t}
   -v_{kj}^{*}(z)\hat{\alpha}_j^\dagger(z) e^{iE_{j}t}
      \right], \nonumber
\label{ansatz}
\end{equation}
where, $\hat{\alpha}_j$ ($\hat{\alpha}_j^\dagger$) are the quasi-particle
annihilation (creation) operators and satisfy Bose commutation 
relations, $u_k$ and $v_k$ are the quasi-particle amplitudes, and $j$ is the 
energy eigenvalue index. We define the operators as common to both the 
species, which is natural and consistent as the dynamics of the species are 
coupled. Furthermore, this reproduces the standard coupled Bogoliubov-de 
Gennes equations at $T=0$ \cite{ticknor_13} and in the limit 
$a_{12}\rightarrow 0$, non-interacting TBEC, the quasi-particle spectra 
separates into two distinct sets: one set for each of the condensates.  From 
the above definitions, we get the following Bogoliubov-de Gennes equations
\begin{subequations}
\begin{eqnarray}
 \hat{{\mathcal L}}_{1}u_{1j}-U_{11}\phi_{1}^{2}v_{1j}+U_{12}\phi_1 \left 
  (\phi_2^{*}u_{2j} -\phi_2v_{2j}\right )&=& E_{j}u_{1j},\;\;\;\;\;\;\\
 \hat{\underline{\mathcal L}}_{1}v_{1j}+U_{11}\phi_{1}^{*2}u_{1j}-U_{12}\phi_1^*\left (
   \phi_2v_{2j}-\phi_2^*u_{2j} \right ) &=& E_{j}v_{1j},\;\;\;\;\;\;\\
 \hat{{\mathcal L}}_{2}u_{2j}-U_{22}\phi_{2}^{2}v_{2j}+U_{12}\phi_2\left ( 
   \phi_1^*u_{1j}-\phi_1v_{1j} \right ) &=& E_{j}u_{2j},\;\;\;\;\;\;\\
 \hat{\underline{\mathcal L}}_{2}v_{2j}+U_{22}\phi_{2}^{*2}u_{2j}-U_{12} \phi_2^*\left ( 
 \phi_1v_{1j}-\phi_1^*u_{1j}\right ) &=& E_{j}v_{2j},\;\;\;\;\;\; 
\end{eqnarray}
\label{bdg}
\end{subequations}
where $\hat{{\mathcal L}}_{1}=
\big(\hat{h}_1+2U_{11}n_{1}+U_{12}n_{2})$, $\hat{{\mathcal L}}_{2}
=\big(\hat{h}_2+2U_{22}n_{2}+U_{12}n_{1}\big)$ and 
$\hat{\underline{\cal L}}_k  = -\hat{\cal L}_k$. 
To solve Eq. (\ref{bdg}) we define $u_k$ and $v_k$'s as linear
combination of $N$ harmonic oscillator eigenstates. 
\begin{eqnarray}
u_{1j} = \sum_{i=0}^N p_{ij}\xi_{i},\;\;
   v_{1j} = \sum_{i=0}^N q_{ij}\xi_{i},\nonumber \\
u_{2j} = \sum_{i=0}^N r_{ij}\xi_{i},\;\;
   v_{2j} = \sum_{i=0}^N s_{ij}\xi_{i},
\label{exp}
\end{eqnarray}
where $\xi_i$ is the $i$th harmonic oscillator eigenstate and
$p_{ij}$, $q_{ij}$,  $r_{ij}$ and $s_{ij}$ are the coefficients of linear
combination. Using this expansion the Eq. (\ref{bdg}) is then
reduced to a matrix eigenvalue equation and solved using standard matrix
diagonalization algorithms.  The matrix has a dimension of $4N\times4N$ and
is non-Hermitian, non-symmetric and may have complex eigenvalues. 
In the present work, to avoid metastable states, we ensure that
$E_j$'s are real during the iteration. The eigenvalue spectrum obtained from 
the diagonalization of the matrix has an equal number of positive and negative
eigenvalues $E_j$'s. The number density $\tilde{n}_k$ of the non-condensate 
atoms is then
\begin{equation}
 \tilde{n}_k=\sum_{j}\{[|u_{kj}|^2+|v_{kj}|^2]N_{0}(E_j)+|v_{kj}|^2\},
 \label{n_k}
\end{equation}
where $\langle\hat{\alpha}_{j}^\dagger\hat{\alpha}_{j}\rangle = (e^{\beta
E_{j}}-1)^{-1}\equiv N_{0}(E_j)$  is the Bose factor of the quasi-particle
state with real and positive energy $E_j$. The coupled 
Eqns. (\ref{gpe}) and(\ref{bdg}) are solved iteratively till the solutions
converge to desired accuracy. However, it should be emphasized 
that, when $T\rightarrow0$, $ N_0(E_j)$'s in Eq. (\ref{n_k}) vanishes.
The non-condensate density is then reduced to 
\begin{equation}
 \tilde{n}_k = \sum_{j}|v_{kj}|^2.
 \label{n_kr}
\end{equation}
Thus, at zero temperature we need to solve the equations self-consistently
as the quantum depletion term $|v_{kj}|^2$ in the above equation is non-zero. 
The contribution from the quantum depletion to the non-condensate is very
small, it is $\approx0.1\%$ for the set of parameters used in our 
calculations. In addition, the solutions to the equations converge in less than
five iterations.
\begin{figure}[h]
 \includegraphics[height=8cm,angle=-90]{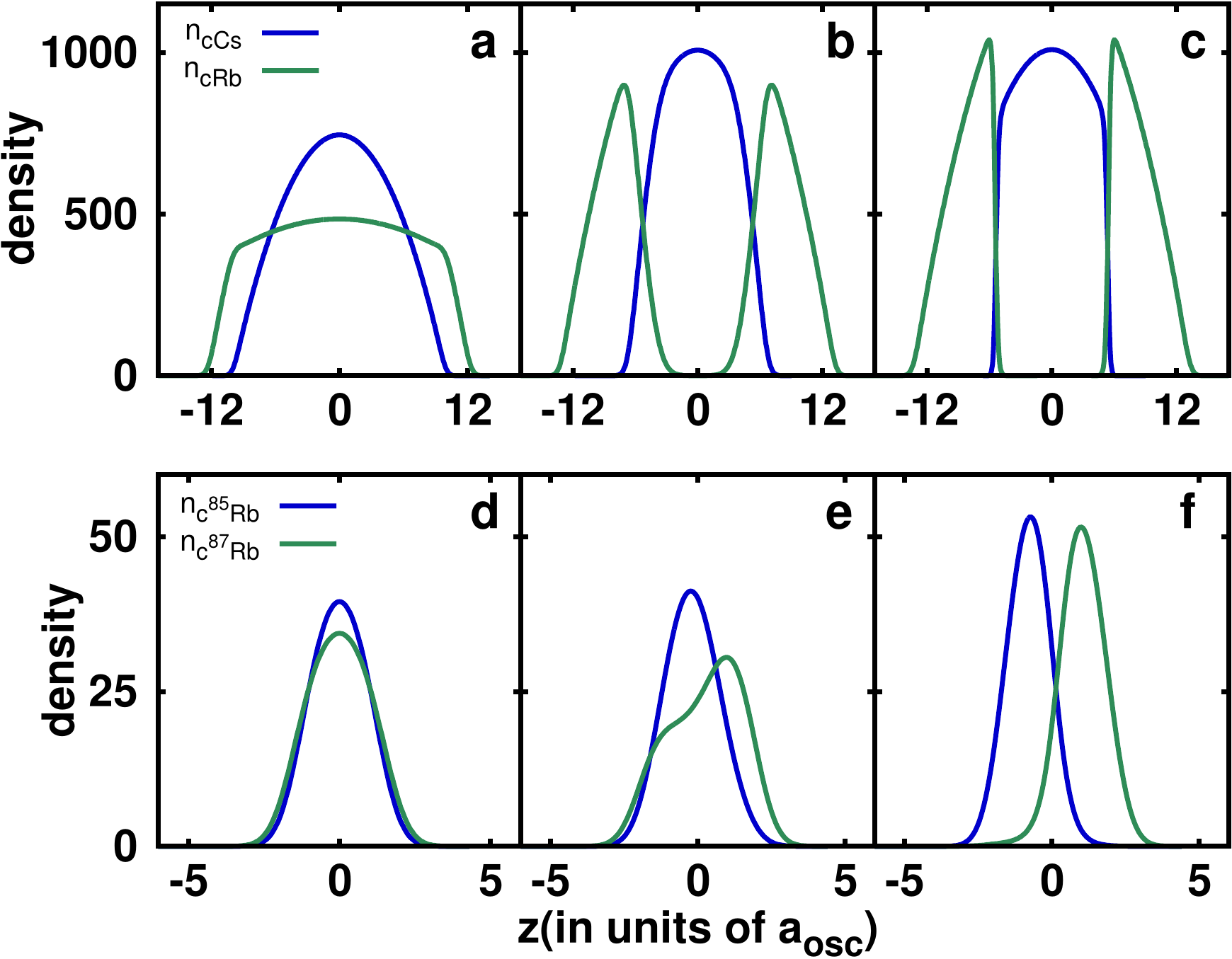}
 \caption{(Color online)
          Transition to phase separation and structure of the density 
          profiles in TBEC. (a-c) show the transition 
          from miscible to {\em sandwich} type density profile with the change 
          in interspecies scattering length $a_{\rm CsRb}$ for a Cs-Rb TBEC
          and correspond to $a_{\rm CsRb} = \{200a_0,310a_0,420a_0\}$ 
          respectively. The density profiles in (c) is referred to as the
          {\em sandwich} type. (d-f) show the transition from miscible to 
          {\em side-by-side} density profile with the change in  
          $a_{^{85}{\rm Rb}^{87}{\rm Rb}}$ for a $^{85}{\rm Rb}-^{87}{\rm Rb}$
          TBEC and correspond to $a_{^{85}{\rm Rb}^{87}{\rm Rb}} 
          = \{100a_0,290a_0,400a_0\}$  respectively. The density profile in 
          (f) is referred to as the {\em side-by-side} type. In the plots
          density is measured in units of $a_{\rm osc}^{-1}$.
         }
 \label{dens_prof}
\end{figure}


\section{Results and discussions}

\subsection{Numerical details }

For the $T=0$ studies we solve the pair of coupled Eqs.~(\ref{gpe}) by 
neglecting the non-condensate density ($\tilde{n}_k=0$) using 
finite-difference methods and in particular, we use the split-step 
Crank-Nicholson method ~\cite{muruganandam_09} adapted for binary 
condensates. The method when implemented with imaginary time propagation is 
appropriate to obtain the stationary ground state wave function of the TBEC. 
Using this solution, and based on Eq.~(\ref{exp}), we cast the Eq. (\ref{bdg}) 
as a matrix eigenvalue equation in the basis of the trapping potential. 
The matrix is then diagonalized using the LAPACK routine
{\tt zgeev} \cite{anderson_99} to find the quasi-particle energies and 
amplitudes, $E_j$, and $u_k$'s and $v_k$'s, respectively. This step is 
the beginning of the first iteration for $T\neq0$ calculations. In which case,
the $u_k$'s and $v_k$'s along with $E_j$ are used to get the initial estimate
of $\tilde{n}_k$ through Eq.~(\ref{n_k}). For this we consider only the
positive energy modes. Using this updated value of $\tilde{n}_k$, the ground 
state wave function of TBEC $\phi_k$ and chemical potential $\mu_k$ are again 
re-calculated from Eq.~(\ref{gpe}). This procedure is repeated till the 
solutions reach desired convergence. In the present work the convergence 
criteria is that the change in $\mu_k$ between iterations should be less 
than $10^{-4}$. In general, the convergence is not smooth and we encounter 
severe oscillations very frequently. To damp the oscillations and accelerate 
convergence we employ successive over (under) relaxation technique for 
updating the condensate (non-condensate) densities\cite{simula_01}. The new 
solutions after ${\rm IC}$ iteration cycle are
\begin{eqnarray}
\phi_{\rm IC}^{\rm new}(z) &= &s^{\rm ov}\phi_{\rm IC}(z) 
                      + (1-s^{\rm ov})\phi_{\rm IC-1}(z),\nonumber \\
\tilde{n}_{\rm IC}^{\rm new}(z)& = &s^{\rm un}\tilde{n}_{\rm IC}(z) 
                      + (1-s^{\rm un})\tilde{n}_{\rm IC-1}(z),
\end{eqnarray}
where $s^{\rm ov}>1$ ($s^{\rm un}<1$) is the over (under) relaxation 
parameter. During the calculation of the $u_k$ and $v_k$, we choose an optimal 
number of the harmonic oscillator basis functions. The conditions based
on which we decide the optimal size are: obtaining reliable Goldstone modes; 
and all eigen values must be real. For the $T=0$ studies
we find that a basis set consisting of 130 harmonic oscillator eigenstates 
is an optimal choice. We observe the Goldstone modes eigenenergies becoming
complex, with a small imaginary component, in the eigen spectrum
when the basis set is very large. So, in the present studies, we ensure 
that there are no complex eigenvalues with an appropriate choice of the 
basis set size.
\begin{figure}[h]
 \includegraphics[height=8cm,angle=-90]{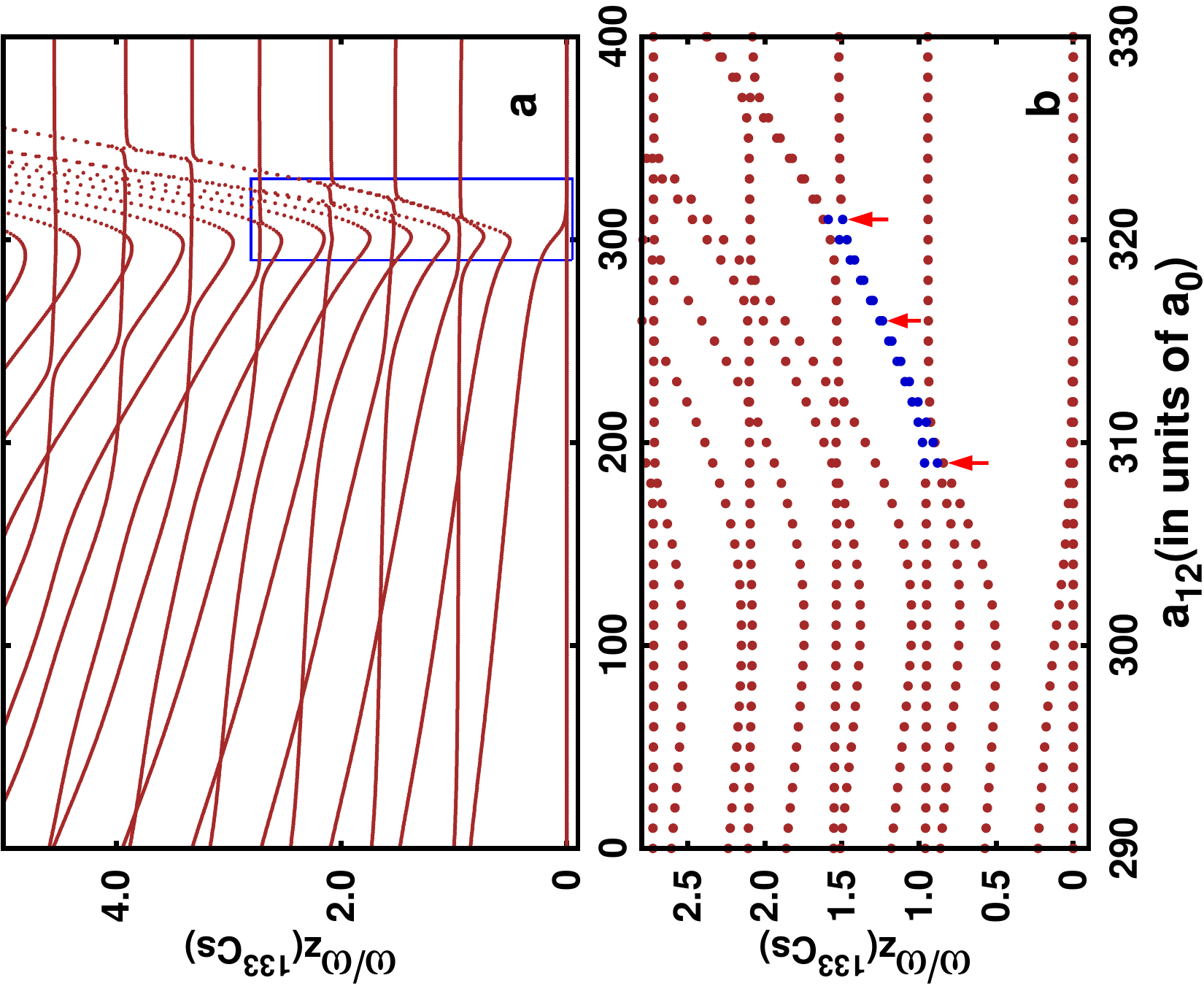}
 \caption{(Color online)
          The evolution of the modes as a function of the interspecies
          scattering length $a_{\rm CsRb}$ in Cs-Rb TBEC. (a) shows the 
          evolution of the low-lying modes  in the domain 
          $0\leqslant a_{\rm CsRb}\leqslant 400 a_0 $ for $N_{^{87}{\rm Rb}} 
          = N_{^{133}{\rm Cs}} = 10^4$. (b) is the enlarged view of the 
          region enclosed within the blue colored rectangular box in figure
          (a) to resolve the avoided crossing and quasi-degeneracy of modes
          (highlighted with dark-blue points). The points marked with 
          red arrows corresponds to interspecies scattering length
          $a_{\rm CsRb} = \{309a_0,316a_0,321a_0\}$ respectively.
          }
         
 \label{mode_evol}
\end{figure}

\begin{figure}[h]
 \includegraphics[height=8cm,angle=-90]{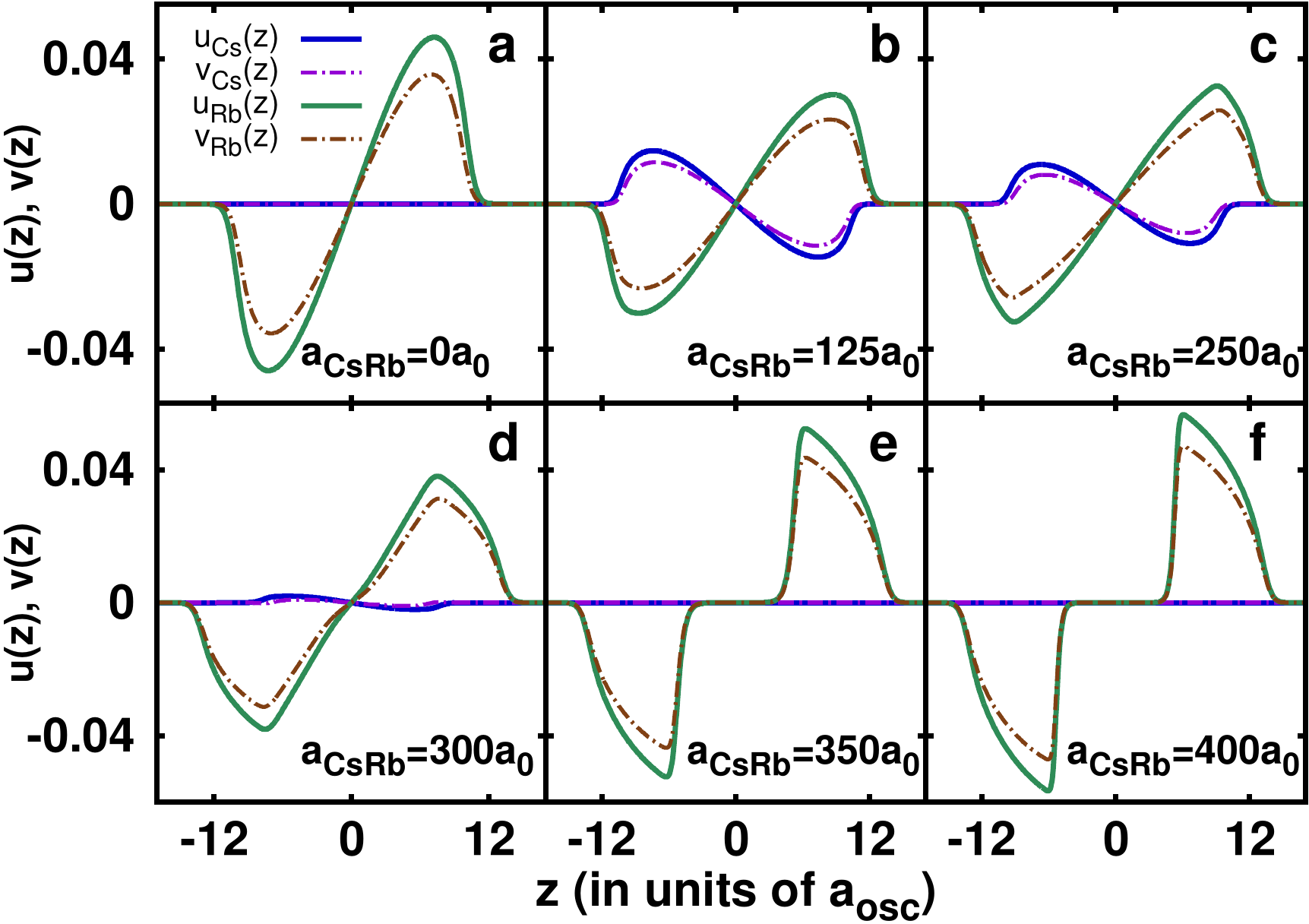}
 \caption{(Color online)
          Evolution of quasi-particle amplitude corresponding
          to the Rb Kohn mode as $a_{_{\rm CsRb}}$ is increased from 
          0 to 400$a_0$. For better visibility $u_{\rm cs}$ and $u_{\rm Rb}$ 
          are scaled by a factor of 1.2. (a) When $a_{_{\rm CsRb}}=0$, it is a 
          Kohn mode of the Rb condensate. (b-d) In the domain 
          $0<a_{_{\rm CsRb}}\lesssim 310a_0$ the mode acquires admixtures from 
          the Cs Kohn mode (nonzero $u_{\rm Cs}$ and $v_{\rm Cs}$). (e-f) At 
          phase separation  $310a_0\lesssim a_{_{\rm CsRb}}$ the mode 
          transforms to a Goldstone mode: $u_{\rm Rb}$ and $v_{\rm Rb}$ have 
          same profile as the $n_{\rm Rb}=|\phi_{\rm Rb}|^2$ but with a phase 
          difference. In the plots $u$'s and $v$'s are in units of 
          $a_{\rm osc}^{-1/2}$.
         }
 \label{3rd_gmode}
\end{figure}


\subsection{Mode evolution of trapped TBEC at $T=0$ }

In TBECs, phase separation occurs when $U_{12}> \sqrt{U_{11}U_{22}}$. For 
the present work, we  consider Cs and Rb as the first and 
second species, respectively. With this identification 
$a_{11}=a_{\rm CsCs} = 280 a_0$ and $a_{22}= a_{\rm RbRb}=100a_0$,
where $a_0$ is the Bohr radius, and arrive at the condition for phase 
separation $a_{12}=a_{_{\rm CsRb}} > 261 a_0$, which is 
smaller than the background value of $a_{_{\rm CsRb}} \approx 650a_0$ 
\cite{lercher_11}. To examine the nature of modes in the neighbourhood of 
phase separation, we compute $E_j$ at $T=0$ and vary 
$a_{_{\rm CsRb}}$, which is experimentally possible with the Rb-Cs Feshbach 
resonance \cite{pilch_09}. The evolution of the low-lying modes in the 
domain $0\leqslant a_{_{\rm CsRb}}\leqslant450 a_0  $ with 
$N_{\rm Rb} = N_{\rm Cs}=10^4$ are computed with 
$\omega_{z({\rm Rb})} = 2\pi\times 3.89 $Hz 
and $\omega_{z({\rm Cs})} = 2\pi\times 4.55 $Hz as in 
ref. \cite{mccarron_11,pattinson_13}. However, to form a quasi-1D system we 
take $\omega_{\perp} = 50 \omega_z$, so that $\hbar\omega_{\perp}\gg \mu_k$.   
For these values, the relevant quasi-1D parameters  
$\alpha = 2a_{_{\rm CsCs}}\sqrt{(\omega_\perp/\omega_z)
(m\omega_\perp/\hbar)} \approx 0.36$ and 
$\gamma = 2(a_{_{\rm CsCs}}/n_{\rm Cs})(m\omega_\perp/\hbar) \approx 10^{-5}$, 
so the system is in the weakly interacting TF regime \cite{petrov_00} and 
mean field description through GP-equation is valid. For this set of 
parameters the ground state is of {\em sandwich} geometry, in which the 
species with the heavier mass is at the center and flanked by the species 
with lighter mass at the edges. An example of the {\em sandwich} profile 
corresponding to the experimentally relevant parameters is shown in 
Fig~(\ref{dens_prof})(c). On the other hand for TBEC with species of equal or 
near equal masses and low number of atoms, in general, the ground
 state geometry is {\em side-by-side}. As an example the {\em side-by-side}
 ground state density profile of $^{85}$Rb-$^{87}$Rb TBEC is shown in
 Fig. ~(\ref{dens_prof})(f). 

From here on we consider the same set of $\omega_z$ ($\omega_{z({\rm Rb})} 
= 2\pi\times 3.89 $Hz and $\omega_{z({\rm Cs})} = 2\pi\times 4.55 $Hz ), 
as mentioned earlier, in the rest of the calculations reported in the 
manuscript. In the computations we scale the spatial and temporal variables 
as $ z/a_{\rm osc( Cs )}$ and $\omega_{z({\rm Cs })}t$ which render the 
equations dimensionless. When $a_{_{\rm CsRb}}=0$, the $U_{\rm CsRb}$ 
dependent terms in Eq.(\ref{bdg}) are zero and the spectrum of the two 
species are independent as the two condensates are decoupled. The system has 
two Goldstone modes, one each for the two species. The two lowest modes with 
nonzero excitation energies are the Kohn modes of the two species, and 
these occur at $ \hbar\omega_{z({\rm Cs})}$ and 
$0.85 \hbar\omega_{z({\rm Cs})}$ for Cs and Rb species, respectively. 
\begin{figure}[h]
\includegraphics[height=8cm,angle=-90]{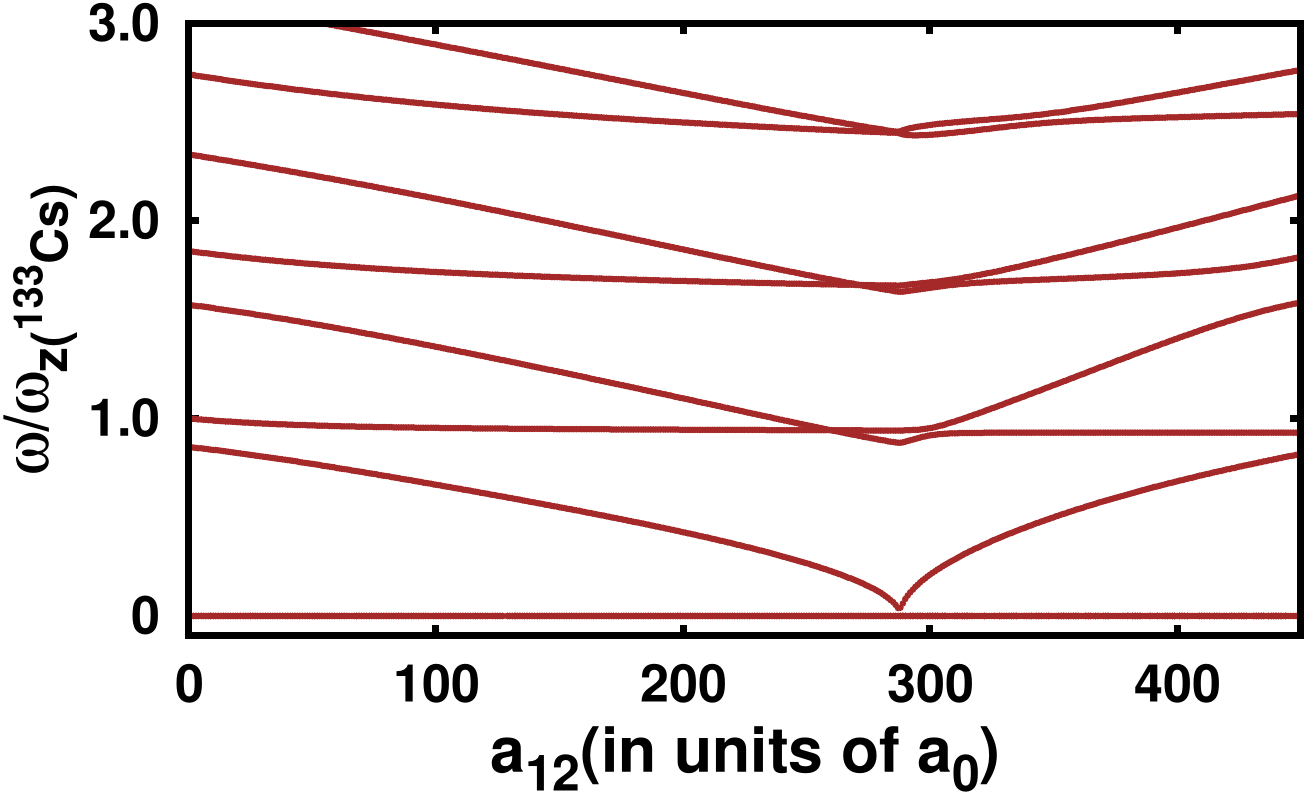}
\caption{(Color online)
          Low-lying modes of  $^{85}$Rb-$^{87}$Rb for
          $N_{^{87}{\rm Rb}} = N_{^{85}{\rm Rb}} = 10^2$ as a function of
          $a_{^{85}{\rm Rb}^{87}{\rm Rb}}$. At phase separation
          the structure of the density profiles is {\em side-by-side}
          and one of the modes goes soft. 
          }
\label{mode_evol1}
\end{figure}


\subsubsection{Third Goldstone mode }

  The clear separation between the modes of the two species is lost 
and mode mixing occurs when $a_{_{\rm CsRb}} > 0$. For example, the Kohn
modes of the two species intermix when $a_{_{\rm CsRb}}>0$, however, there is 
a difference in the evolution of the mode energies. The energy of the Rb Kohn 
mode decreases, but the one corresponding to Cs remains steady at 
$\hbar\omega_{z({\rm Cs})}$. At higher $a_{_{\rm CsRb}}$ the energy of 
the Rb Kohn mode decreases further and goes soft at phase separation 
($U_{\rm CsRb}> \sqrt{U_{\rm CsCs}U_{\rm RbRb}}$)
when $a_{_{\rm CsRb}}\approx 310a_0$. This introduces a new Goldstone mode of
the Rb BEC to the excitation spectrum. The reason is, for the parameters 
chosen, the density profiles at phase separation {\color{black}assume
{\em sandwich} geometry} with Cs BEC at the center and Rb BEC at the edges. 
So, the Rb BECs at the edges are effectively two topologically distinct BECs 
and there are two Goldstone modes with the same $|u_{\rm Rb}|$ and 
$|v_{\rm Rb}|$ but different phases. A similar result of the Kohn mode going 
soft was observed for single species BEC confined in a double well 
potential \cite{salasnich_99}. Although, the the two systems are widely 
different, there is a common genesis to the softening of the Kohn mode, and 
that is the partition of the one condensate cloud into two distinct ones. This 
could be, in our case by another condensate or by a potential barrier as 
in ref. \cite{salasnich_99}. 

 To examine the mode evolution with the experimentally realized parameters
\cite{mccarron_11}, we repeat the computations with 
$\omega_{\perp ({\rm Cs})}=2\pi\times 40.2 $Hz and 
$\omega_{\perp ({\rm Rb})}=2\pi\times 32.2 $Hz. With these parameters the 
system is not strictly quasi-1D as $\hbar\omega_{\perp k}\approx \mu_k$ for
$N_{\rm Cs}=N_{\rm Rb}=10^4$, however, as $\omega_{z k}\ll \omega_{\perp k}$ 
there must be qualitative similarities to a quasi-1D system \cite{egorov_13}.
Indeed, with the variation of $a_{_{\rm CsRb}}$ the modes evolve similar to the 
case of $\omega_{\perp k}=50\omega_{zk}$ and low-lying $\omega$s 
are shown in Fig. \ref{mode_evol}(a). The evolution of the Rb Kohn mode
functions ($u_{\rm Rb}$ and $v_{\rm Rb}$) with $a_{_{\rm CsRb}}$ are shown 
in Fig. \ref{3rd_gmode}. It is evident that when $a_{_{\rm CsRb}}=0$
(Fig. \ref{3rd_gmode}(a)), there is no
admixture from the Cs Kohn mode ( $u_{\rm Cs}=v_{\rm Cs}=0$). However, when 
$0<a_{_{\rm CsRb}}\lesssim 310a_0$ the admixture from the Cs Kohn mode 
increases initially and then goes to zero as we approach 
$U_{\rm CsRb}> \sqrt{U_{\rm CsCs}U_{\rm RbRb}}$ (Fig. \ref{3rd_gmode}(b-f) ). 

  One striking result is, the Rb Kohn mode after going soft at 
$a_{_{\rm CsRb}}\approx 310a_0$, as shown in Fig. \ref{mode_evol}(a), 
continues as the third Goldstone mode for $310a_0 < a_{_{\rm CsRb}}$. This is 
different from the evolution of the zero energy mode in TBEC with
{\em side-by-side} density profiles.  In this case after phase
separation, $z$-parity symmetry of the system is broken
and the zero energy mode regains energy. So, there are only two 
Goldstone modes in the system. This is evident from Fig. \ref{mode_evol1}, 
where we show the mode evolution of $^{85}$Rb-$^{87}$Rb mixture with 
{\em side-by-side} density profiles at phase separation. The parameters of the 
system considered are $N_{^{85} {\rm Rb}}=N_{^{87} {\rm Rb}} = 10^2$ with the 
same $\omega_{zk} $ and $\omega_{\perp k}$ as in the Rb-Cs mixture. Here,
we use intra-species scattering lengths as $99a_0$ and  $100a_0 $ for 
$^{85}$Rb and $^{87}$Rb, respectively and tune the inter-species interaction 
for better comparison with the Rb-Cs results. This is, however, different 
from the experimental realization \cite{papp_08}, where the intra-species 
interaction of $^{85}$Rb is varied. A similar result was reported in an 
earlier work on quasi-2D system of TBEC \cite{ticknor_13}. 
\begin{figure}[h]
 \includegraphics[height=8.5cm,angle=-90]{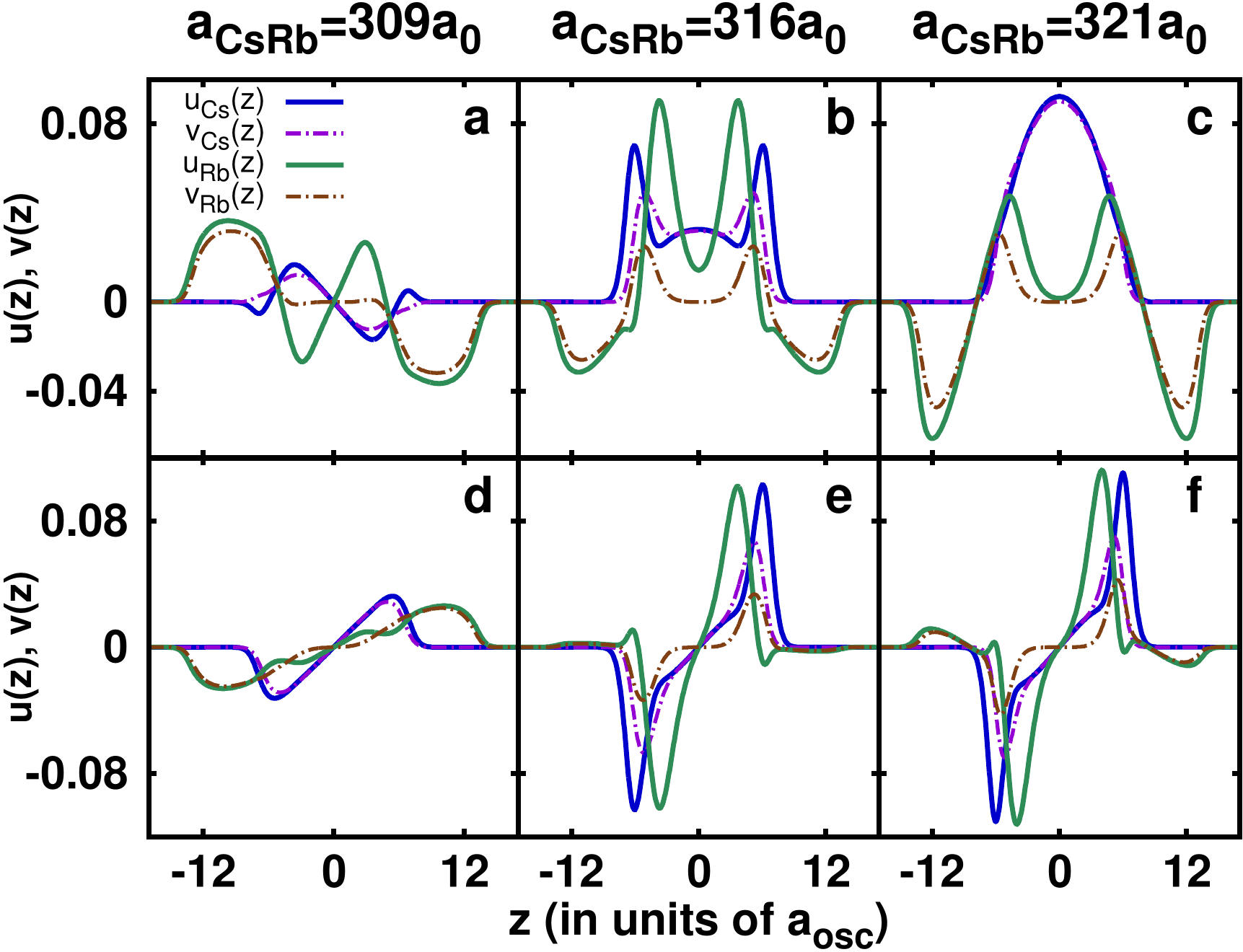}
 \caption{(Color online)
          The quasi-particle amplitudes of the 5th and 6th modes at 
          quasi-degeneracy. (a-c) The quasi-particle amplitudes $u_k$'s and 
          $v_k$'s of the $5^{\rm th}$ mode for 3 values of $a_{_{\rm CsRb}}$
          represented and marked by blue points and  red arrows, respectively,
          in Fig. (\ref{mode_evol}). (d-f) The quasi-particle 
          amplitudes $u_k$'s and $v_k$'s corresponding to the $6^{\rm th}$
          mode for the same values of $a_{_{\rm CsRb}}$. In the plots
          $u_k$'s and $v_k$'s are in the units of $a_{\rm osc}^{-1/2}$.
         }
 \label{56_modes}
\end{figure}


\subsubsection{Avoided crossings and quasi-degeneracy }

From Fig. \ref{mode_evol}(a), it is evident that there are several
instances of avoided level crossings as $a_{_{\rm CsRb}}$ is varied to higher
values. These arise from the changes in the profile of $n_{ck}(z)$, the 
condensate densities, as the $u_k$ and $v_k$ depend on $n_{ck}(z)$ through 
the BdG equations. For this reason, the number of avoided crossings is high 
around the critical value of $a_{_{\rm CsRb}} $, where there  is a significant 
change in the structure of $n_{ck}(z)$ due to phase separation. Another 
remarkable feature which emerges when $a_{_{\rm CsRb}}  > 310a_0$ are the 
avoided crossings involving three modes. As an example, the mode evolution 
around one such case involving the Kohn mode is shown in 
Fig. \ref{mode_evol}(b). Let us, in particular, examine the 5th
and 6th modes, the corresponding mode energies in the domain of interest
($309 a_0 \leqslant a_{_{\rm CsRb}} \leqslant 321a_0$) are represented by blue
colored points in Fig. \ref{mode_evol}(b). At $a_{_{\rm CsRb}} = 309a_0$, the 
6th mode is the Kohn mode, which is evident from the dipolar structure of 
the $u_k$ and $v_k$ as shown in Fig. \ref{56_modes}(d). The closest approach
of the three modes, 4th, 5th and 6th, occurs when 
$a_{_{\rm CsRb}}\approx 311 a_0$, at this point the 4th mode is transformed 
into Kohn mode. For $a_{_{\rm CsRb}} > 311a_0$, the 5th and 6th mode energies 
are quasi degenerate and pushed to higher values.  For example, at 
$a_{_{\rm CsRb}} = 316 a_0$ the energies of the 5th and 6th modes
are 1.24$\hbar\omega_{z{\rm (Cs)}}$ and 1.25$\hbar\omega_{z{\rm (Cs)}}$,
respectively. However, as shown in Fig. \ref{56_modes}(b) and (e), the
structure of the corresponding $u_k$ and $v_k$ show significant difference. It
is evident that for the 5th mode $u_{\rm Cs}$ and $u_{\rm Rb}$ correspond
to principal quantum number $n$ equal to 0 and 2, respectively. On the
other hand, for the 6th mode both $u_{\rm Cs}$ and $u_{\rm Rb}$ have
$n$ equal to 1. At $a_{_{\rm CsRb}}\approx 320 a_0$, the two modes (5th and 
6th) undergo their second avoided crossing with a third mode, the 7th mode.
After wards, for $a_{_{\rm CsRb}} > 320 a_0$, the 5th mode remains steady at
1.50$\hbar\omega_{z{\rm (Cs)}}$, and the 6th and 7th are quasi degenerate.
To show the transformation of the 5th and 6th modes beyond the second
avoided crossing, the $u_k$ and $v_k$ of the modes are shown in
Fig. \ref{56_modes}(c) and (f) for $a_{_{\rm CsRb}} = 321 a_0$. It is evident 
from the figures that the $u_{\rm Cs}$ and $v_{\rm Cs}$ of the 5th mode
undergoes a significant change in the structure: the central dip at
$a_{_{\rm CsRb}}   < 321 a_0$, visible in Fig \ref{56_modes}(b), is modified 
to a maxima.
\begin{figure}[h]
 \includegraphics[height=8cm,angle=-90]{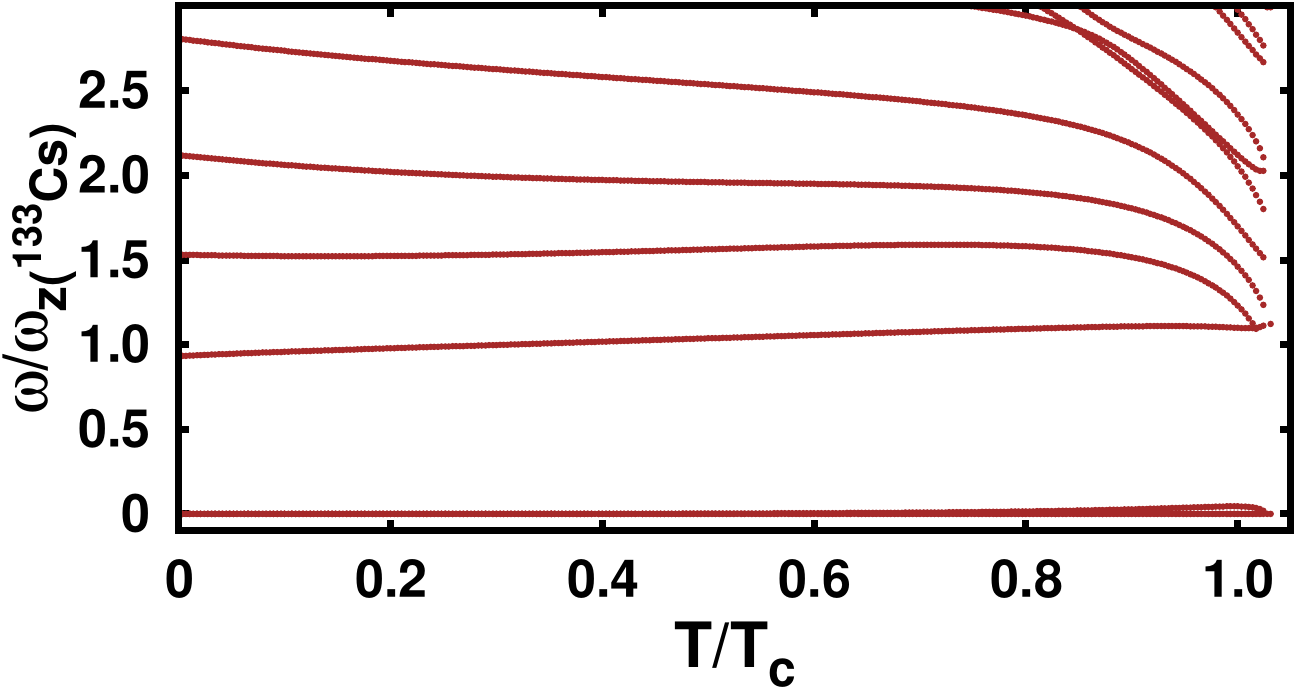}
 \caption{(Color online)
          Frequencies ($\omega_j$) of the low-lying modes at $T/T_c\neq 0$. The
          filled circles (brown) are the excitation energies from the HFB-Popov
          theory with $N_{\rm Rb}=N_{\rm Cs}=10^3$. The blue colored points
          indicate the location of the mode bifurcation.
         }
 \label{mode_fin}
\end{figure}


\subsection{Mode evolution of trapped TBEC at $T\neq0$}

For the $T\neq0$ calculations, as 
mentioned earlier, we solve the coupled Eq. (\ref{gpe}) and (\ref{bdg}) 
iteratively till convergence. After each iteration, $\phi_k(z)$ are 
renormalized so that 
\begin{equation}
  \int_{-\infty}^{\infty} \left [ |\phi_k(z)|^2  + \tilde{n}_k(z) \right]dz 
       = N_k,
\end{equation}
where $k$ is either Rb or Cs. To improve convergence, we use successive over 
relaxation, but at higher $T$ we face difficulties and require careful choice 
of the relaxation parameters. For computations, we again consider the trap 
parameters
$\omega_{\perp ({\rm Cs})}=2\pi\times 40.2 $Hz and 
$\omega_{\perp ({\rm Rb})}=2\pi\times 32.2 $Hz with coinciding trap centers,
the number of atoms as $N_{\rm Rb} = N_{\rm Cs}=10^3$ and 
$a_{\rm CsRb} = 650a_0$. The evolution of $\omega$ (mode frequency) with 
$T$ is shown in Fig. \ref{mode_fin}, where the $T$ is in 
units of $T_c$, the critical temperature of ideal bosons in quasi-1D harmonic 
traps defined through the relation 
$N=(k_{\rm B}T_c/\hbar\omega_z) \ln (2k_{\rm B}T_c/\hbar\omega_z)$ 
\cite{ketterle_96}, where $N$ is the number of atoms.
Considering that $\omega_{z({\rm Rb})}  < \omega_{z({\rm Cs})}$, the 
critical temperature of Rb is lower than that of Cs. So, for better
description we scale the temperature with respect to the $T_c$ of Rb atoms,
and here after by $T_c$ we mean the critical temperature of Rb atoms.
From the figure, when $T/T_c\geqslant 0.2$ the 
Kohn mode energy increases with $T/T_c$. This is consistent with an earlier 
work on HFB-Popov studies in single species condensate \cite{gies_04}, but
different from the trend observed in ref. \cite{hutchinson_97,dodd_98}. 
The increase in Kohn mode energy could arise from 
an important factor associated with the thermal atoms. In the HFB-Popov 
formalism the collective modes oscillates in a static thermal cloud background 
and dynamics of $\tilde{n}_k$ is not taken into account. In TBECs the effects 
of dynamics of $\tilde{n}_k$  may be larger as $\tilde{n}_k$ is large at the 
interface. An inclusion of the full dynamics of the thermal cloud in the 
theory would ensure the Kohn mode energy to be constant at all 
temperatures \cite{hutchinson_2000}. The Goldstone modes, on the other hand, 
remain steady\cite{gies_04}.

 The trend in the evolution of the modes indicates bifurcations at
$T/T_c \approx 1$ and is consistent with the theoretical observations in
single species condensates \cite{hutchinson_97,dodd_98,gies_04}.  At
this temperature, as evident from Fig. \ref{mode_fin}, the Kohn mode and the 
mode above it (which has principal quantum no $n=2$ for both the species ) 
merges. For brevity, the location 
of the mode bifurcation is indicated by the blue colored points in 
Fig. \ref{mode_fin}. This is one of the bifurcations emerging from the Rb 
atoms crossing the critical temperature, above this temperature there are 
no Rb condensates atoms. At $ T > T_c$ the Cs condensate density is still 
non-zero as Cs has higher critical temperature. So, there may be another 
mode-bifurcation at the critical temperature of Cs. A reliable calculation for 
this would, however, require treating the interaction between thermal Rb atoms
and Cs condensate more precisely. For this reason in the present work we do
not explore temperature much higher than the $T_c$ of Rb atoms and  the 
possibility of the second mode bifurcation shall be examined in our future 
works. In the case of single species calculations, at $T/T_c > 1 $ 
the mode frequencies coalesce to the mode frequencies of the trapping 
potential. In the present work we limit the calculations to 
$0\leqslant T/T_c \leqslant 1.1$, so that $T/T_c \ll T_d/T_c$. Here, 
$T_d\approx (N_{\rm Rb} + N_{\rm Cs})\hbar\omega_z/k_{\rm B}$ is the 
degeneracy temperature of the system and in the present case
$T_d\approx 437$nK. The results for $T/T_c > 0.65$ may have
significant errors as the HFB-Popov theory gives accurate results at 
$T/T_c\leqslant 0.65$ \cite{dodd_98}. We have, however, extended the 
calculations to $T/T_c > 0.65$ like in Ref. \cite{hutchinson_97} to study
the mode bifurcation.
\begin{figure}[h]
 \includegraphics[height=8cm,angle=-90]{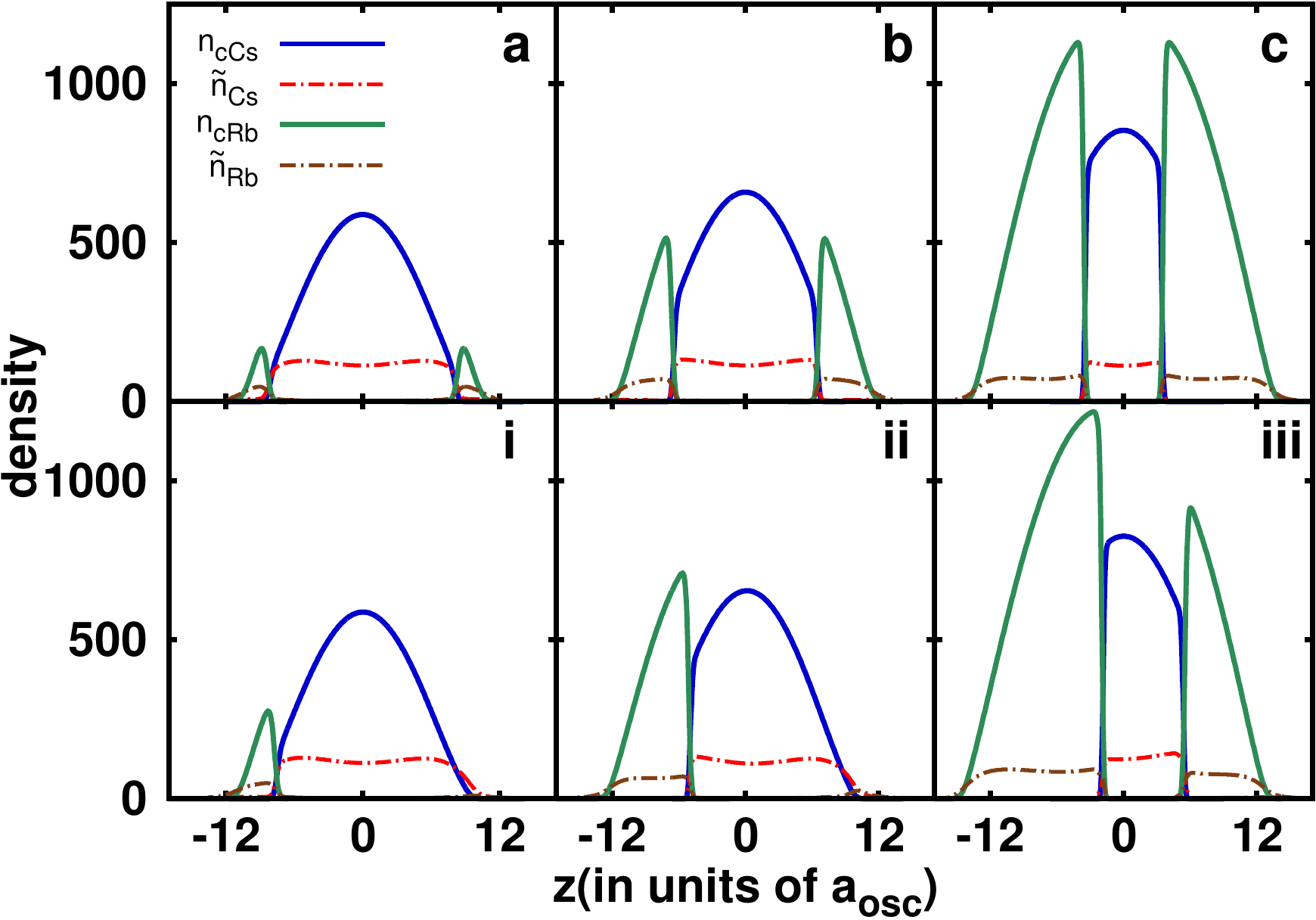}
 \caption{(Color online)
          Density profile of $n_c$ and $\tilde{n}$ at 25nK. (a), (b) 
          and (c) correspond to $N_{\rm Rb} = 840 (N_{\rm Cs}=8570)$,
          $N_{\rm Rb}=3680 (N_{\rm Cs}=8510)$, and 
          $N_{\rm Rb}=15100 (N_{\rm Cs}=6470)$, respectively, with 
          coincident trap centers. (i), (ii) and (iii) correspond to same 
          atom numbers as the previous sequence, however, the trap centers 
          are shifted relatively by 0.8$a_{\rm osc (Cs)}$.
          In the plots density is measured in units of $a_{\rm osc}^{-1}$.
         }
 \label{den_prof}
\end{figure}

 To examine the profiles of $n_{ck}$ and $\tilde{n}_k$, we compute the 
densities at 25nK for three cases, these are 
$N_{\rm Rb} = 840 (N_{\rm Cs}=8570)$,
$N_{\rm Rb}=3680 (N_{\rm Cs}=8510)$, and $N_{\rm Rb}=15100 (N_{\rm Cs}=6470)$.
The same set was used in the previous work of Pattinson, et al. at 
$T=0$ \cite{pattinson_13} and correspond to three regimes considered 
($N_{\rm Cs} > N_{\rm Rb}$,  $N_{\rm Cs} \approx N_{\rm Rb} $, and
$N_{\rm Cs} < N_{\rm Rb}$) in the experimental work of McCarron, 
et al. \cite{mccarron_11}. Consider the trap centers, along $z$-axis, are
coincident, then  $\tilde{n}_k$ and $n_{ck}$ are symmetric about $z=0$, 
and are shown in Fig. \ref{den_prof}(a-c). In all the cases,
$n_{\rm Cs}$ is at the center. This configuration is
energetically preferred as heavier atomic species at the center has smaller 
trapping potential energy and lowers the total energy. In the experiments, 
the trap centers are not exactly coincident. So, to replicate the experimental 
situation we shift the trap centers, along $z$-axis, by 0.8$a_{\rm osc (Cs)}$ 
and $n$ are shown in Fig. \ref{den_prof}(i-iii). For 
$N_{\rm Rb} = 840 (N_{\rm Cs}=8570)$ and $N_{\rm Rb}=3680 (N_{\rm Cs}=8510)$, 
Fig. \ref{den_prof}(i-ii), the $n_{ck}$ and $\tilde{n}_k$ are
located sideways. So, there are only two Goldstone modes in the excitation
spectrum. But, for $N_{\rm Rb}=15100 (N_{\rm Cs}=6470)$, 
Fig. \ref{den_prof} (iii), $n_{\rm Cs}$ is at the center with $n_{\rm Rb}$ at 
the edges forming {\em sandwich} geometry and hence has three 
Goldstone modes. In all the cases $\tilde{n}_k$ have
maxima in the neighbourhood of the interface and the respective $n_{ck}$s are 
not negligible. So, we can expect larger $n_{ck}$-$\tilde{n}_k$ coupling in 
TBECs than single species condensates. For the 
$N_{\rm Rb}=3680 (N_{\rm Cs}=8510)$ and
$N_{\rm Rb}=15100 (N_{\rm Cs}=6470)$ cases, $n_{ck}$ 
are very similar to the results of 3D calculations at $T=0$ 
\cite{pattinson_13}. However, it requires a 3D calculation 
to reproduce $n_{ck}$ for $N_{\rm Rb} = 840 (N_{\rm Cs}=8570)$ as the
relative shift $\delta x$ is crucial in this case.


\section{Conclusions}
TBECs with strong inter-species repulsion with {\em sandwich } 
density profile at phase-separation are equivalent to three coupled 
condensate fragments. Because of this we observe three Goldstone modes in the 
system after phase-separation. At higher inter-species interactions, we 
predict avoided crossings involving three modes and followed with the 
coalescence or quasi-degeneracy of two of the participating modes. 
At $T\neq 0$ there are mode bifurcations close to the $T/T_c\approx 1 $.


\begin{acknowledgments}
We thank K. Suthar and S. Chattopadhyay for useful 
discussions. The results presented in the paper are based on the computations 
using the 3TFLOP HPC Cluster at Physical Research Laboratory, Ahmedabad, India.
We also thank the anonymous referees for their thorough review and 
valuable comments, which contributed to improving the quality of the 
manuscript.
\end{acknowledgments}

\bibliography{bec}{}
\bibliographystyle{apsrev4-1}

\end{document}